# Remote sensing of pressure inside deformable microchannels using light scattering in Scotch tape

KyungDuk Kim[1], HyeonSeung Yu[1], Joonyoung Koh[2], Jung H. Shin[2], Wonhee Lee[2], and YongKeun Park[1,*]

[1]Department of Physics, KAIST, 291 Daehakro, Daejeon 305-701, Republic of Korea
[2]Graduate School of Nanoscience and Technology, KAIST, 291 Daehakro, Daejeon 305-701, Republic of Korea
*Corresponding author: yk.park@kaist.ac.kr

**We present a simple but effective method to measure the pressure inside a deformable micro-channel using laser scattering in a translucent Scotch tape. Our idea exploits the fact that the speckle pattern generated by a turbid layer is sensitive to the changes in an optical wavefront of an impinging beam. A change in the internal pressure of a channel deforms the elastic channel, which can be detected by measuring speckle patterns of a coherent laser that has passed through the channel and the Scotch tape. We demonstrate that internal pressure can be remotely sensed with the resolution below 0.1 kPa within a pressure range of 3 kPa after calibration. With its high sensitivity, reproducibility, and easy applicability, the present method will find direct and diverse applications.**

Gas flows in microchannels are one of the important components for the efficient transport of matter or energy in micro-electro-mechanical systems (MEMS) [1]. Over the past decades, a variety of gaseous systems has been miniaturized into microchannel systems, such as a cooling system [2] and a gas analyzer [3]. Gas dynamics in microchannels is also fundamentally interesting physical phenomenon since rarefied gas dynamics arises from the tiny scale [4]. Various experimental and theoretical studies have been performed [5, 6] to effectively utilize the gas inside micro-fabricated channels.

In the study of gas flows in microchannels, pressure plays an important role in characterizing or controlling the gaseous environment inside microchannels. For engineering the gas flow inside the channel, monitoring pressure should be accompanied [7]. One of the approaches for direct pressure sensing is to create a hole at the channel and insert a needle-like tip of pressure gauge, but it dramatically affects the pressure. To avoid direct sensing, the external transducers [8] and the microfabricated built-in sensors [9] were proposed. However, those methods are cost-ineffective in consideration of their low resolutions [10].

Alternatively, various optical methods to measure pressure have been developed. One type of the methods is based on optical fiber sensors, such as Mach-Zehnder interferometry [11], Febry-Perot etalon [12], and Fiber Bragg grating [13, 14]. They can provide high sensitivity of detecting the pressure-induced deformation in order of wavelength. However, the use of fiber sensors with microchannels is usually restricted due to the difficulty in the integration with the channel of a small size. In order to gauge pressure inside deformable channels non-invasively, a diffraction grating embedded in channels [15] and the interferometric measurements of channel membrane [16] were also proposed. Unfortunately, it is necessary to apply complicated external measurement schemes for these methods.

Herein, we propose a simple optical method to measure the pressure of gas inside a deformable microchannel. This method exploits speckle patterns formed by a scattering layer, which scatters a coherent laser that has passed through the transparent channel whose internal pressure is to be measured. A minute distortion in the optical wavefront of a laser beam, associated with the deformation of a channel due to the internal pressure, can be converted into speckle patterns via light interference intensity patterns. High sensitivity is gained due to the randomness of the geometric structure of a scattering layer, enabling the change in internal pressure to be easily detected by an ordinary image sensor.

Our idea exploits the scattering of light from a medium of disordered configuration. Random scattering occurs when light encounters a medium with inhomogeneous distribution of refractive indexes, or turbid medium. Then, light experiences seemingly stochastic but deterministic scattering events, and the optical phases in a resultant optical field would be highly scrambled. Thus, with a coherent illumination, transmitted light interferes and forms a complex speckle pattern. Due to the complexity of scattering, tiny differences in the incident light are conveyed and amplified in the transmitted field [17].

Historically, speckle patterns have been utilized for precise optical sensing of physical properties of scattering media or rough surfaces, including displacement [18], roughness [19], velocity [20], temperature [21], and solution concentration [22]. Recently, speckle patterns have also been exploited for measuring the optical properties of scattered light such as incident angles [23], wavelengths [24-26], and images [27, 28]. To our best knowledge, however, there had been no investigation for sensing pressure using laser speckle.

The principle of the present method is illustrated in Fig. 1. To measure the pressure inside a deformable channel, we impinged a collimated laser beam on the surface of the transparent microchannel, below which a layer of Scotch tape is inserted as a scattering medium. The optical phase delay of the beam transmitting the channel is determined by the geometric shape and refractive index of the channel. In general, however, this phase delay is small so that this difference in phase is barely observable by a conventional imaging system. Here, a scattering layer, having a rough surface, further scrambles and diffuses the light to form random speckle patterns via interference [Fig. 1(a)].

The interior wall of a deformable channel expands when the internal pressure inside the channel increases. The phase of the transmitted beam thus now contains the information relevant to the internal pressure inside the channel [Fig. 1(b)]. The scattering layer let the difference in wavefronts be reflected in the output scattered field. Therefore, a minute and invisible change in the shape of channels due to the pressure change can be translated into an apparent change in the speckle field, which can be detected by a simple image sensor.

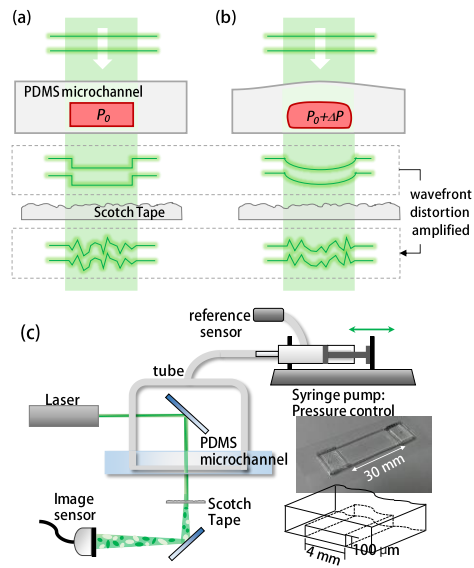

Fig. 1. (a) Schematic diagram that describes the change in wavefront of laser beam which passed through a transparent channel and a scattering layer. (b) Deformation of wavefront caused by increase of internal pressure, and role of scattering layer that amplifies the wavefront difference. (c) Experimental setup. Pressure inside the microchannel was controlled by a syringe pump.

In order to demonstrate proof of these concepts, we implemented a setup as illustrated in Fig. 1(c). It consisted of the parts for controlling and measuring the pressure inside a channel, and the optical part to generate and detect speckle patterns from a scattering layer. We controlled the pressure using a syringe (Kovax-Syringe 30 ml, Korea Vaccine Co., Republic of Korea), whose volume was regulated with a syringe pump (PHD ULTRA CP 4400, Harvard Apparatus, USA) by the infusion and withdrawal of the syringe. To monitor the internal pressure directly as a reference, we directly connected a pressure sensor (PG-30, Copal Electronics, USA) to the syringe body with a tube with an inner diameter of 1/16 inch (Teflon PFA tubing, Dupont, USA). The resolution of the reference pressure sensor is 0.01 kPa, and the maximum limit of measurable pressure difference is 3.5 kPa. We used microchannels made with a transparent and deformable material, PDMS (Polydimethylsiloxane), and the channel has dimensions of 30 mm (length) × 4 mm (width) × 100 μm (height), as shown in the inset of Fig. 1(c). The two ends of the microchannel were connected with tubes with an inner diameter of 1/16 inch (Teflon PFA tubing, Dupont, USA). The two ends of the tubes were then connected by a T-shaped-connector, to avoid the pressure gradient inside the channel. At the other end of the connector, the syringe was connected so that the pressure inside microchannel could be controlled with the syringe pump.

For a coherent illumination source, we used a diode-pumped solid state laser ($\lambda$ = 532 nm, 50 mW, Shanghai Dream Laser Co., Shanghai, China). The laser beam was incident on the surface of a PDMS microchannel. A scattering layer, made by attaching a single layer of a Scotch tape (Scotch Magic Tape, 3M, USA) on a slide glass, was located under the channel with a separation of 2 cm. Then the light that transmits through the channel will diffuse out by the scattering layer. Then the speckle patterns were recorded by a CCD image sensor (INFINITYlite, Lumenera, USA). The distance between the scattering layer and the CCD was adjusted so that one speckle spot contains approximately 10 × 10 pixels in order to maximize the signal-to-noise ratio.

To demonstrate the capability of the present method, we systemically measured speckle patterns at various internal pressures that were controlled by the syringe pump and monitored by a reference pressure sensor [Fig. 2]. The pressure was initially set as atmospheric pressure, and monotonically increased and then decreased until it returned to the initial pressure with a constant rate of 0.09 kPa/s. At the same time, the speckle intensity patterns of the transmitted beams were recorded by the CCD sensor. Figures 2(a)-(h) present the representative speckle patterns at various internal pressures in the absence and the presence of the Scotch tape layer.

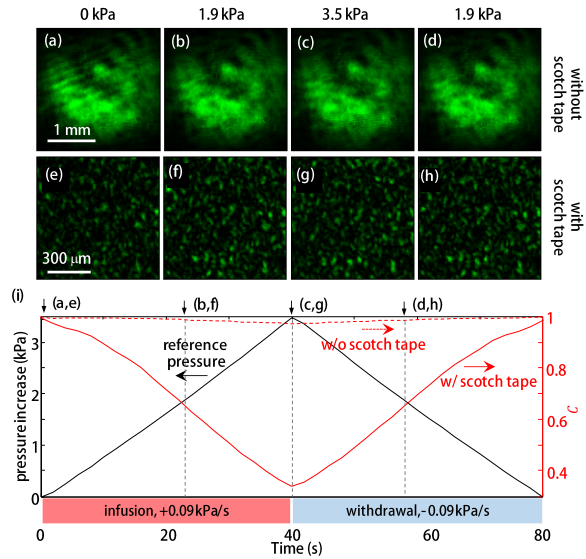

Fig. 2. Recorded output field at several values of pressure in a microchannel, (a-d) with and (e-h) without scotch tape. Speckle pattern is observed when laser beam is scattered by scotch tape. (i) Change of correlation coefficient when the pressure was increased and then decreased.

In the absence of the Scotch tape, the images show beam profiles with little diffraction patterns caused by the channel structure. These intensity images do not show significant difference between different pressures [Figs. 2(a)-(d)]. When we inserted a layer of a Scotch tape beneath the microchannel, speckle patterns appeared and their patterns significantly and sensitively changed at different pressures [Figs. 2(e)-(h)].

In order to quantitatively analyze the change in speckle patterns, we calculated the correlation coefficient $C$ between two pixelated two-dimensional images of equal size, $I_1$ and $I_2$, defined as follows:

$$C = \frac{\sum_{m,n}(I_1(m,n)-\overline{I_1})(I_2(m,n)-\overline{I_2})}{\sqrt{\left(\sum_{m,n}(I_1(m,n)-\overline{I_1})^2\right)\left(\sum_{m,n}(I_2(m,n)-\overline{I_2})^2\right)}}, \quad (1)$$

where $I(m,n)$ is the intensity at the $m^{th}$ row and $n^{th}$ column of $I$, $\overline{I}$ is its mean value, and the range of summations is the entire image. Figure 2(i) shows the correlation coefficients as a function of time. The correlation coefficient at a given pressure was calculated with the initial image at 0 kPa. The reference pressure (the black line) exhibits a little asymmetric part near the peak due to the backlash of a syringe pump.

Both the correlation coefficients with and without the Scotch tape follow the trends of a pressure change: they fall off when the pressure increases, and rise again as the pressure decreases. Notably, with the Scotch tape, the correlation coefficient changes more sensitively with the pressure change, as compared to the case without the Scotch tape. At 3.5 kPa, the maximum pressure in this measurement, the correlation coefficient with the Scotch tape drops to 0.34, while it was 0.97 without the Scotch tape. This result clearly shows that a layer of Scotch tape causes light diffusion that results in the effective conversion of wavefront change to the intensity change via light interference. Furthermore, the symmetry of the pressure-correlation curves in Fig. 2(i) implies that, at arbitrary pressure, a speckle image would show a similar pattern regardless of whether the pressure is increasing or decreasing.

In order to demonstrate the capability of the present method as a practical sensor, we addressed the repeatability and hysteresis of the pressure sensing scheme. In Fig. 3(a), five repeated measurements were performed in which internal pressure was increased from 0 to 3 kPa. Each measurement of $C$ presents a highly reproducible result. To quantitatively address the reproducibility, repeatability at a specific pressure was calculated as the ratio of the largest error of $C$ to the averaged $C$ at the given pressure. The averaged repeatability was calculated as 1.5% at the range of P less than 1 kPa, and it increases as $P$ becomes larger due to the decrease in the value of $C$.

To study the hysteresis of the present method, we compared the $P$-$C$ curves between increasing and decreasing pressure. Figure 3(b) shows that the present method exhibits slight hysteresis; $P$-$C$ curves for increasing $P$ (the red lines) show $C$ values higher than those for decreasing $P$ (the black lines). We attribute this hysteresis in $C$ measurements to the hysteresis of the channel deformation.

To further investigate the performance of the present method, we quantified the resolution and hysteresis at various measurement time intervals between consecutive acquirements of speckle images. In order to quantitatively compare the hysteresis, we found the pressures when $C$ becomes 0.5 while increasing and decreasing pressures [Fig. 4(a)]. We defined the hysteresis as the difference in pressure for the increasing and decreasing states [inset, Fig. 4(a)]. In addition, we defined the resolution as the maximum error range in pressure at arbitrary pressure, reflecting both hysteresis and repeatability. Figure 4(b) shows the hysteresis calculated at various time intervals. As the measurement time interval increases, the hysteresis tends to be smaller. This is due to the viscoelastic property of PDMS, which requires some amount of time to restore its original shape after deformation.

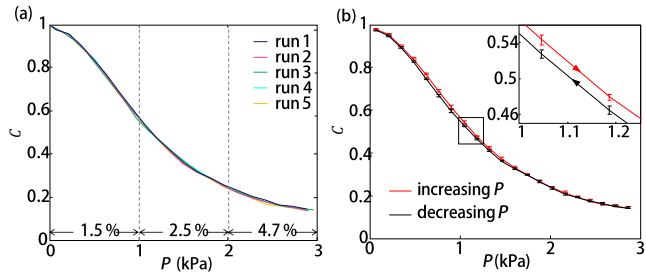

Fig. 3. (a) Correlation curves of repeated measurements, increasing pressure from 0 to 3 kPa. (b) Hysteresis curve with error bars of repeated measurements. Inset represents the expansion of rectangle box in the original figure.

Figure 4(c) shows the averaged resolution ranging from 0.05 kPa to 0.12 kPa. Resolution also becomes smaller as the measurement time interval increases, mainly due to the decrease of hysteresis. Compared to the reported resolution of existing optical fiber sensors for pressure such as 0.11 kPa for Fiber Bragg Grating type [29] and 0.69 kPa for Febry-Pérot type [30], the present method achieved a better resolution with a significantly simpler and more cost-effective setup.

To further investigate the effect of different types of scattering layers to the performance of the present method, we tested multi-layered Scotch tape and 15° diffuser [Fig. 4(d)]. As expected, the $P$-$C$ curve with single Scotch tape layer shows significant enhancement in performance, compared to the case without Scotch tape. Interestingly, the increase of the number of tape layers does not show a significant enhancement in sensitivity in pressure measurement. The $p$-value between groups of measurements obtained with one layer of Scotch tape and ten layers of Scotch tapes was greater than 0.4. There was even no difference when the type of scattering layer is replaced from Scotch tape to 15° diffuser. The p-value between measurements for one layer of Scotch tape and 15° diffuser was greater than 0.5, proving no significant difference among the uses of various types of scattering layers. These results imply that the correlation curve of scattering sensor is robust to the choice of scattering layers and one can use any type of scattering layer for sensor design.

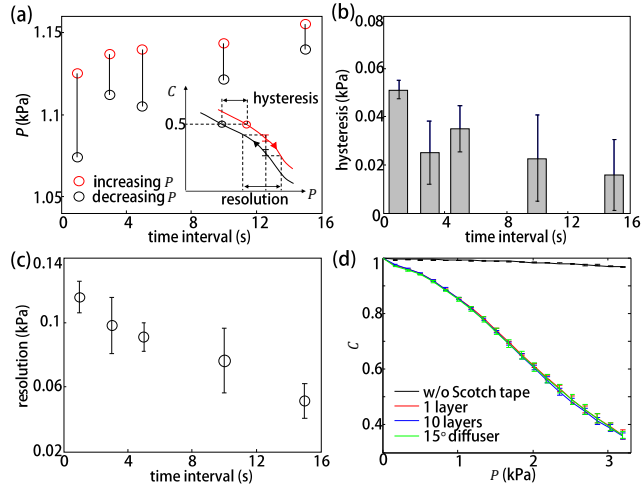

Fig. 4. (a) The change of pressure at which the correlation coefficient between speckle images is halved due to the change in time interval between consecutive measurements. Inset represents the definition of hysteresis and resolution. The change of (b) hysteresis and (c) resolution when the time interval between consecutive measurements was changed. (d) Comparison between correlation curves with different thickness and type of a diffusive layer.

In this letter, we presented a novel optical technique to measure the pressure inside a deformable microchannel. Exploiting light scattering resulting from translucent Scotch tape, a small change in optical wavefront due to internal pressure can result in a significant change in speckle patterns, allowing the detection of the internal pressure using a cost-effective but precise method. This is analogous to the concept of phase contrast microscopy, which converts a small change in the wavefront due to a phase object into a significant change in the intensity pattern via interference using a phase mask [31]. In previous experiments dealing with light transport through complex media, small phase changes should be avoided and minimized [32, 33]; however, this work rather utilizes this phenomenon for effective pressure sensing.

The present method can be readily implemented to existing microchannel settings because the addition of a Scotch tape layer and pressure calibration are all that are required. Measuring speckle patterns with a simple image sensor, such as a smartphone camera can provide a sensitivity higher than optical fiber sensor techniques that requires a light source and a spectrometer, which is bulky and expensive. In addition, this method enables remote in-situ sensing of pressure, which may find scenarios in which samples, such as reactive or toxic samples, should not contact the sensor. Furthermore, the present method allows pressure measurements at various points of channel by simply illuminating a laser beam to the target point. One limitation of this method is that it needs a calibration process for each experimental setting because the nonlinear $P$-$C$ curves highly depend on the deformability modulus of a channel. However, this limitation can be overcome using a channel with standardized elasticity.

Moreover, another popular application of microchannels is micro fluidic system. The proposed scheme can also be extended to measure the pressure of liquid inside microfluidics. Potential further requirement may involve the calibration of various refractive indices of liquid, because it affect the retardation of wavefront and resultant speckle. Also, the existence of liquid causes the lensing effect of an inflated channel [34]. However, we expect that the proposed sensor can be implemented in liquid with proper adjustment of system parameters.

Our technique guarantees highly reproducible measurement of gas pressure inside a deformable microchannel. With the demonstrated simple implementation and high sensitivity, we envision the present method having direct and diverse applications in which measuring pressure is important

**Funding.** This work was supported by the National Research Foundation (NRF) of Korea (2014K1A3A1A09063027, 2013M3C1A3063046, 2012-M3C1A1-048860, 2014M3C1A3052537), and APCTP.